\documentclass[conference]{IEEEtran}
\IEEEoverridecommandlockouts

\usepackage{cite}
\usepackage{amsmath,amssymb,amsfonts}
\usepackage{algorithmic}
\usepackage{graphicx}
\usepackage{textcomp}
\usepackage{xcolor}
\usepackage{amsmath}
\usepackage{amssymb}
\usepackage{bm,braket}
\usepackage{float} 
\usepackage{tikz} 
\usepackage{verbatim}
\usetikzlibrary{quantikz} 
\usepackage{booktabs}
\usepackage{tabularx}
\usepackage{ragged2e}
\usepackage{xcolor}

\usepackage{hyperref}
\usepackage{multirow}

\hypersetup{
     colorlinks   = true,
     citecolor    = blue
}

\def\be{\begin{equation}}
\def\ee{\end{equation}}
\def\bea{\begin{eqnarray}}
\def\eea{\end{eqnarray}}
\def\bes{\begin{eqnarray}}
\def\ees{\end{eqnarray}}
\def\bi{\begin{itemize}}
\def\ei{\end{itemize}} % ------- Define Greek Lowercase ------

\def\BibTeX{{\rm B\kern-.05em{\sc i\kern-.025em b}\kern-.08em
    T\kern-.1667em\lower.7ex\hbox{E}\kern-.125emX}}

\makeatletter
\newcommand{\linebreakand}{
\end{@IEEEauthorhalign}
\hfill\mbox{}\par
\mbox{}\hfill\begin{@IEEEauthorhalign}
}
\makeatother
    
\begin{document}

\title{Quantum Enhanced Deep Reinforcement Learning for Distribution Network Reconfiguration  
\thanks{*These authors contributed equally to this work. Research funded by NIST Grant GONANB24D219-1 ``Empowering the Future Grid: Quantum Training for Software Innovation \& Utility Applications.`` We thank Shaked Regev and Phil Lotshaw for their insights throughout this study. }
}

\author{\IEEEauthorblockN{Daniel Germain*}
\IEEEauthorblockA{\textit{EPB Quantum} \\
Chattanooga, TN, USA \\
germaind@epbqn.net}
\and
\IEEEauthorblockN{Mason Blanchard*}
\IEEEauthorblockA{\textit{EPB Quantum} \\
Chattanooga, TN, USA \\
blanchardm@epbqn.net}
\and
\IEEEauthorblockN{Audrey Versteegen*}
\IEEEauthorblockA{\textit{EPB Quantum} \\
Chattanooga, TN, USA \\
versteegena@epbqn.net}
\and
\IEEEauthorblockN{Nora Bauer}
\IEEEauthorblockA{
\textit{IonQ Inc.}\\
Chattanooga, TN, USA \\
nora.bauer@ionq.co}
\linebreakand
\IEEEauthorblockN{Jonathan Mei}
\IEEEauthorblockA{
\textit{IonQ Inc.}\\
Chattanooga, TN, USA \\
jmei@ionq.co}
\and
\IEEEauthorblockN{Claudio Girotto}
\IEEEauthorblockA{
\textit{IonQ Inc.}\\
Chattanooga, TN, USA \\
claudio.girotto@ionq.co}
\and
\IEEEauthorblockN{Paul Smith}
\IEEEauthorblockA{
\textit{EPB Quantum}\\
Chattanooga, TN, USA \\
smithpj@epb.net}
\and
\IEEEauthorblockN{Martin Roetteler}
\IEEEauthorblockA{
\textit{IonQ Inc.}\\
College Park, MD, USA \\
martin.roetteler@ionq.co}
}
\maketitle

\begin{abstract}
Modern smart grids generate increasingly rich operational data and deploy greater numbers of remotely controlled switching assets than ever before, creating new opportunities for data-driven distribution network reconfiguration (DNR). DNR optimizes grid topology to improve efficiency, reduce power losses, and maintain operational constraints, but finding optimal switch configurations remains a computationally demanding combinatorial problem. In this paper, we propose a deep reinforcement learning (DRL) framework for static DNR to systematically investigate how the Q-network architecture affects both solution quality and learning performance. We evaluate multiple architectures within a controlled DRL framework, varying only the Q-network to isolate the contribution of each architectural choice. Our results show that quantum-enhanced Q-networks can achieve higher rewards with fewer parameters than their classical counterpart, suggesting that such architectures are a promising direction for combinatorial power systems optimization.  
\end{abstract}


\section{Introduction}
Electrical distribution networks deliver electricity from transmission systems to end users. Due to their wide geographic spread and the resistive losses associated with power delivery, a significant portion of generated energy is dissipated before reaching consumers. To mitigate these losses, utilities have explored various strategies, with distribution network reconfiguration (DNR) emerging as an effective and economical solution \cite{civanlar1988feeder,baranwu1989reconfiguration}. DNR involves selectively opening and closing network switches to modify the topology of the distribution system while maintaining operational constraints. Through appropriate reconfiguration, utilities can reduce power losses, improve voltage profiles, and enhance overall system reliability with minimal additional infrastructure~\cite{lotfi2024dnr}.

Despite these advantages, DNR presents a challenging optimization problem. The underlying AC power-flow equations are nonlinear, while practical distribution networks are typically required to maintain a radial topology. Consequently, DNR is formulated as a mixed-integer nonlinear optimization problem and is generally NP-hard \cite{bienstock2019nphard,lavorato2012radiality}. Classical optimization approaches include exhaustive and branch exchange searches \cite{civanlar1988feeder,baranwu1989reconfiguration,morton2000bruteforce}, metaheuristics such as genetic algorithms \cite{nara1992ga}, and mixed-integer convex reformulations \cite{lavorato2012radiality}, among others \cite{lotfi2024dnr}. These approaches can therefore become computationally expensive as network size increases, particularly when the optimal configuration must be repeatedly determined under changing operating conditions~\cite{skolfield2020operations}. In recent years, the increasing deployment of smart meters and grid-monitoring infrastructure has created opportunities for data-driven approaches that can learn reconfiguration strategies directly from system states~\cite{Schuetz_2022}.

A variety of machine learning approaches have been investigated for DNR, including artificial neural networks (ANNs) such as convolutional neural networks (CNNs) and long short-term memory (LSTM) networks \cite{zhang2018dlreview}. These methods have been used to learn relationships between network states, operating conditions, and desirable reconfiguration strategies \cite{Gholizadeh2023,zhang2018dlreview}. For example, CNN-based approaches have been developed to learn relationships between network topology and voltage stability, while LSTM models have been used to learn mappings between load distributions and optimal reconfiguration strategies~\cite{Gholizadeh2023}. Although these approaches can provide rapid inference after training, supervised learning generally requires labeled examples of optimal configurations and may have difficulty adapting when operating conditions differ from those represented in the training data. Furthermore, the combinatorial nature of the reconfiguration problem can make it difficult for conventional learning approaches to directly model the sequential decision process involved in selecting a series of switching actions.

Reinforcement learning (RL) provides a natural framework for this sequential decision-making problem \cite{sutton2018rl}. Rather than learning directly from a dataset of precomputed optimal configurations, an RL agent learns a policy by interacting with the distribution network environment and receiving rewards based on the resulting system state. This enables the agent to learn reconfiguration strategies without requiring an explicit mapping between every possible network state and its optimal configuration. RL is particularly attractive for DNR because the agent can repeatedly evaluate candidate switching actions and learn to optimize objectives such as power loss while satisfying operational and topological constraints. Moreover, once trained, the learned policy can generate decisions rapidly, making RL well suited to environments in which operating conditions vary over time~\cite{Gholizadeh2023,oh2020online}.

Deep reinforcement learning (DRL), which combines reinforcement learning with neural networks to approximate the action-selection policy or value function \cite{mnih2015dqn,watkins1992qlearning}, has demonstrated success in reducing power losses in practical distribution networks~\cite{Di2020}. Deep Q-network (DQN) \cite{mnih2015dqn} and NoisyNet-DQN \cite{fortunato2018noisynet} approaches have subsequently been applied to DNR, with NoisyNet introducing learnable parameterized noise to promote exploration during training~\cite{Wang2021}. Subsequent work has extended DQN-based approaches through techniques such as prioritized experience replay \cite{schaul2016per} and modified reward formulations, while also considering objectives beyond loss minimization, including reliability improvement~\cite{Gholizadeh2023,Gautam2023}. These results demonstrate the potential of DRL to learn effective reconfiguration policies without explicitly solving the underlying combinatorial optimization problem at every operating point \cite{duan2020voltagecontrol}.

In parallel, quantum computing has been proposed as a route to solve hard power system optimization tasks \cite{eskandarpour2020gridprimer,golestan2023qcpower}. Within machine learning specifically, quantum reinforcement learning (QRL) has emerged as a promising research direction, investigating whether quantum circuits can be incorporated into reinforcement learning architectures to improve representation, parameter efficiency, or learning performance \cite{meyer2024survey}. Approaches including quantum policy gradients \cite{jerbi2021policies}, variational quantum policies \cite{Skolik2022}, and quantum Q-networks \cite{Chen2020VQC,Lockwood2020} have been investigated across a range of reinforcement learning problems. However, much of the existing QRL literature has focused on small-scale benchmark environments \cite{Skolik2022,Lockwood2020,Chen2020VQC}, leaving the performance of quantum reinforcement learning on larger, practically motivated optimization problems less well established.

In an effort to scale quantum reinforcement learning to problems with higher-dimensional state spaces, hybrid quantum-classical Q-networks have been proposed with varying levels of success~\cite{chen2024deep,pmlr-v148-lockwood21a}. In this work, we propose a hybrid quantum-classical Q-network for distribution network reconfiguration and evaluate its ability to minimize line losses in the IEEE 33 and 69 bus systems. In simulation, the proposed quantum reinforcement learning framework successfully identifies the known optimal network configurations for both systems. For the IEEE 33 bus system, a multi-seed ablation study is conducted to investigate the relationship between model size, training dynamics, and solution quality, demonstrating that the quantum models can achieve higher mean final rewards while requiring fewer trainable parameters than larger classical Q-networks. Finally, trained 12-qubit models for both the IEEE 33 and 69 bus systems are executed on the IonQ Forte Enterprise-1 trapped-ion quantum processor, with their hardware inference outputs compared against the corresponding noiseless-simulation results to quantify the impact of hardware noise on model performance.

The remainder of this paper is organized as follows. Section \ref{sec:background} provides background on distribution network reconfiguration, deep reinforcement learning, and quantum reinforcement learning. Section \ref{sec:drl_algo} describes the proposed DRL environment and its formulation for network reconfiguration. Section \ref{sec:q_networks} presents the Q-network architectures considered in this study, and section \ref{sec:results} summarizes the experimental results. Lastly, section \ref{sec:conclusion} concludes with key findings and directions for future work.

\section{Background}\label{sec:background}
\subsection{Reinforcement Learning}

\begin{figure*}[t]
\centering
\includegraphics[width=0.85\textwidth]{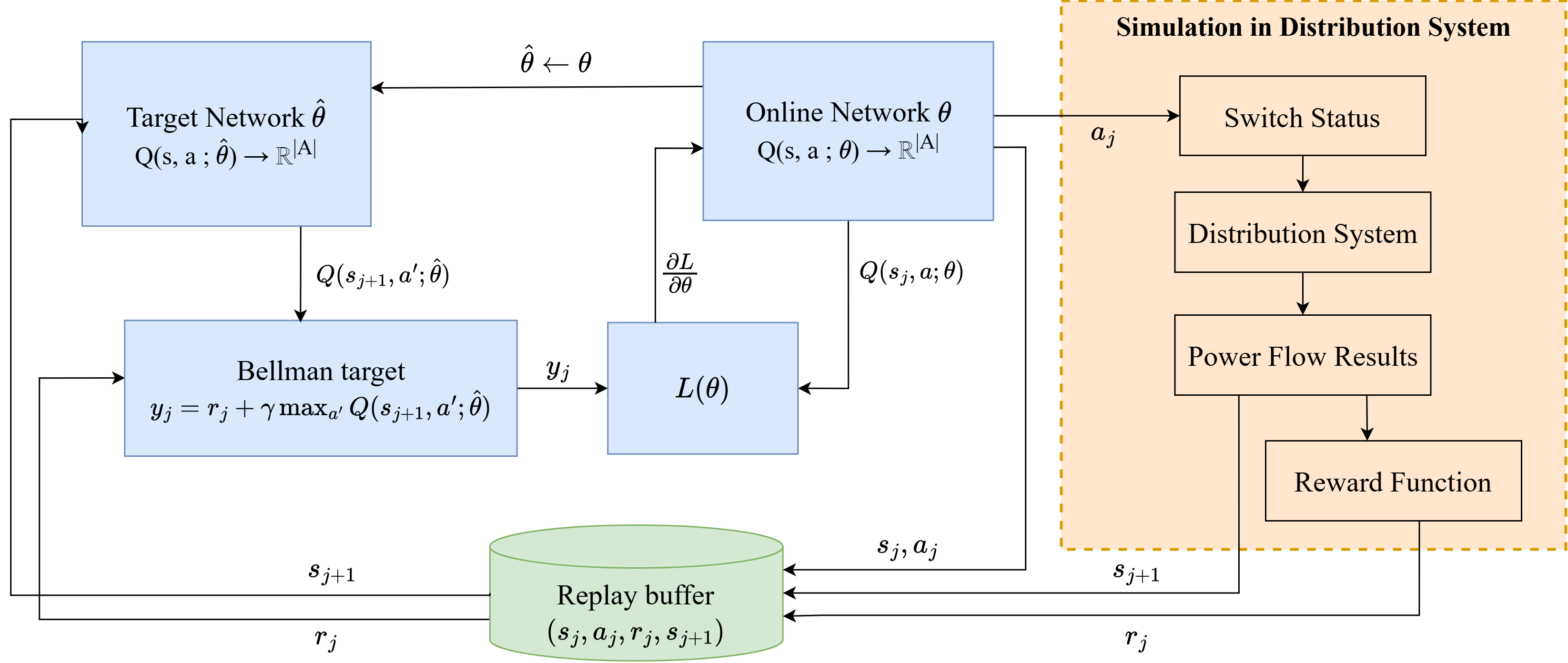}
\caption{The deep reinforcement learning training procedure used in this work. 
Transitions are generated through interaction with a simulated distribution network and are stored in the replay buffer. The replay buffer is sampled to train the online 
Q-network. The online and target networks are used to construct the Bellman 
target, which is compared with the predicted action value to compute the 
training loss.}
\label{fig:drl_training}
\end{figure*}

Reinforcement learning (RL) is a machine learning framework for sequential decision-making in which an agent learns through interactions with an environment $E$~\cite{sutton2018rl}. At time step $t$, the agent observes the environment state $s_t \in S$, selects an action $a_t \in A$ from the set of available actions, receives a reward signal $r_t$, and the environment transitions to a subsequent state $s_{t+1}$. This interaction continues until a terminal condition, such as a time limit or terminal state, is reached.

The RL problem can be formalized as a Markov decision process (MDP), defined by the tuple $(S,A,P,R,\gamma)$~\cite{sutton2018rl}. Here, $S$ denotes the set of possible environment states, $A$ the set of possible actions, and $P(s_{t+1}\mid s_t,a_t)$ the probability of transitioning to state $s_{t+1}$ after taking action $a_t$ in state $s_t$. The reward function $R(s_t,a_t)$ specifies the immediate reward associated with a state-action pair, while $\gamma\in[0,1]$ is the discount factor controlling the relative importance of future rewards. The discounted return from time step $t$ is defined as

\begin{equation}
R_t = \sum_{t'=t}^{T} \gamma^{t'-t} r_{t'},
\end{equation}
where $T$ denotes the time step at which the episode terminates. The objective of the agent is to learn a policy $\pi$ that maximizes the expected return. The action-value function under policy $\pi$ is therefore defined as

\begin{equation}
Q^\pi(s,a)
=
\mathbb{E}_{\pi}\left[
R_t \mid s_t=s,a_t=a
\right],
\end{equation}
where $Q^\pi(s,a)$ represents the expected discounted return obtained by taking action $a$ in state $s$ and subsequently following policy $\pi$.

\subsection{Q-Learning}

RL algorithms can broadly be categorized as model-based or model-free~\cite{sutton2018rl}. Model-based methods use or learn a representation of the environment dynamics, such as the transition probabilities $P(s_{t+1}\mid s_t,a_t)$ and a reward model, which can then be used for planning. In contrast, model-free methods learn a policy or value function directly from interactions with the environment without explicitly learning a model of its transition dynamics. This approach is well suited to the DNR formulation considered in this work because the state resulting from a switching action can be directly evaluated through power-flow analysis, while the corresponding reward can be explicitly calculated from the resulting electrical and topological characteristics. Consequently, the agent does not need to learn an approximation of either the system transition dynamics or the reward function and can instead directly learn the long-term value of its switching actions. Q-learning is a model-free RL algorithm that accomplishes this by learning the value associated with each state-action pair~\cite{watkins1992qlearning}.

In Q-learning, the quality of taking action $a$ in state $s$ is represented by the action-value function $Q(s,a)$. The optimal action-value function $Q^*(s,a)$ represents the maximum expected discounted return obtainable from a state-action pair and satisfies the Bellman optimality equation,

\begin{equation}
Q^*(s,a)=
\mathbb{E}\left[
r_t+\gamma\max_{a'}Q^*(s_{t+1},a')
\mid s_t=s,a_t=a
\right].
\label{eq:bellman}
\end{equation}
The Bellman equation expresses the value of a state-action pair recursively as the immediate reward plus the discounted value of the best action available from the subsequent state. During Q-learning, this relationship is used to construct the one-step Bellman target

\begin{equation}
y_t =
r_t + \gamma \max_{a'} Q(s_{t+1},a'),
\label{eq:bellman_target}
\end{equation}
where $y_t$ provides a target estimate for the current state-action value $Q(s_t,a_t)$. The difference between the Bellman target and the current estimate,

\begin{equation}
\delta_t =
y_t-Q(s_t,a_t),
\label{eq:td_error}
\end{equation}
is known as the temporal-difference (TD) error. In tabular Q-learning, the action-value estimate is iteratively updated toward the Bellman target according to

\begin{equation}
Q(s_t,a_t)
\leftarrow
Q(s_t,a_t)
+
\alpha
\left[
y_t-Q(s_t,a_t)
\right],
\label{eq:q_update}
\end{equation}
where $\alpha$ is the learning rate controlling the magnitude of each update. Through repeated interactions with the environment, these updates progressively improve the action-value estimates toward the optimal action-value function $Q^*$.

Once the action values have been learned, the corresponding greedy policy selects the action with the highest estimated value,

\begin{equation}
a_t =
\underset{a\in\mathcal{A}}{\arg\max}\,
Q(s_t,a).
\label{eq:greedy_policy}
\end{equation}
Although the greedy policy in Eq.~\ref{eq:greedy_policy} exploits the action currently estimated to be optimal, relying exclusively on this strategy during training can prevent the agent from discovering potentially better actions. To balance exploitation of the current value estimates with exploration of alternative actions, Q-learning commonly employs an $\epsilon$-greedy action-selection strategy. At each time step, the agent selects a random action with probability $\epsilon$ and the action with the highest estimated value with probability $1-\epsilon$,

\begin{equation}
a_t =
\begin{cases}
\text{random action}, & \text{with probability } \epsilon, \\[4pt]
\underset{a\in\mathcal{A}}{\arg\max}\; Q(s_t,a), & \text{with probability } 1-\epsilon.
\end{cases}
\label{eq:epsilon_greedy}
\end{equation}
The parameter $\epsilon$ therefore controls the exploration-exploitation trade-off. During training, $\epsilon$ can be gradually decreased so that the agent initially explores a wider range of actions before increasingly exploiting the learned action-value estimates. Once training is complete, the learned policy can be executed greedily by selecting the action with the maximum estimated value.

\subsection{Deep Q-Learning}

Traditional action-value methods often use a tabular representation of $Q(s,a)$, requiring a separate value for each state-action pair. This approach becomes impractical as the state space grows, particularly for environments with high-dimensional or continuous state representations. Deep Q-learning (DQL) addresses this limitation by approximating the action-value function using a neural network known as a Deep Q-Network (DQN)~\cite{mnih2015dqn}.

\begin{equation}
Q(s,a;\theta)\approx Q^*(s,a),
\end{equation}
where $\theta$ represents the trainable parameters of the DQN. The network receives the current state as input and outputs an estimated action value for each available action. During training, actions are selected according to the $\epsilon$-greedy policy described in Eq.~\ref{eq:epsilon_greedy}.

The Bellman target introduced in Eq.~\ref{eq:bellman_target} provides the basis for training the DQN. Rather than directly updating individual entries of a Q-table toward the Bellman target, deep Q-learning adjusts the network parameters $\theta$ such that the predicted value $Q(s_t,a_t;\theta)$ approaches the corresponding target value. However, using a neural network to approximate the action-value function introduces additional instability because the value estimates used to construct the training targets change as the network parameters are updated. DQL therefore employs a separate target network, parameterized by $\theta^{-}$, to provide more stable estimates when constructing the Bellman targets~\cite{mnih2015dqn}.

In a standard DQN, the target network is used both to select and evaluate the maximizing action at the subsequent state. Double Deep Q-Network (Double DQN) instead separates these two operations to reduce the overestimation of action values~\cite{vanhasselt2016deep}. The online Q-network is used to select the action with the highest estimated value at the subsequent state, while the target network evaluates the value of the selected action. Given a transition $(s_t,a_t,r_t,s_{t+1})$, the Double DQN target is therefore

\begin{equation}
y_t =
r_t+
\gamma
Q\left(
s_{t+1},
\underset{a'}{\arg\max}\,Q(s_{t+1},a';\theta);
\theta^{-}
\right).
\label{eq:double_dqn_target}
\end{equation}
Here, the online Q-network with parameters $\theta$ selects the maximizing action, while the target network with parameters $\theta^{-}$ provides the value estimate for that action. The Double DQN target in Eq.~\ref{eq:double_dqn_target} therefore serves the same role as the Bellman target in tabular Q-learning: it provides the value toward which the current action-value estimate is updated. The difference is that, rather than directly modifying a stored value $Q(s_t,a_t)$, the prediction is updated indirectly by optimizing the parameters of the Q-network.

The online Q-network is trained by minimizing the mean squared error (MSE) between its predicted action value and the Double DQN target,

\begin{equation}
\mathcal{L}(\theta)
=
\frac{1}{B}
\sum_{i=1}^{B}
\left[
y_i -
Q(s_i,a_i;\theta)
\right]^2
\label{eq:dqn_loss}
\end{equation}
where $B$ is the minibatch size and $y_i$ is the target value computed using Eq.~\ref{eq:double_dqn_target}. Minimizing this loss moves the online network's action-value estimates toward their corresponding Bellman targets. Only the online-network parameters $\theta$ are updated through gradient descent, while the target-network parameters $\theta^{-}$ are held fixed between target-network updates and are periodically updated from the online network. The overall training procedure is illustrated in Fig.~\ref{fig:drl_training}. 

\begin{figure*}[h] 
    \centering
\includegraphics[width=0.85\linewidth]{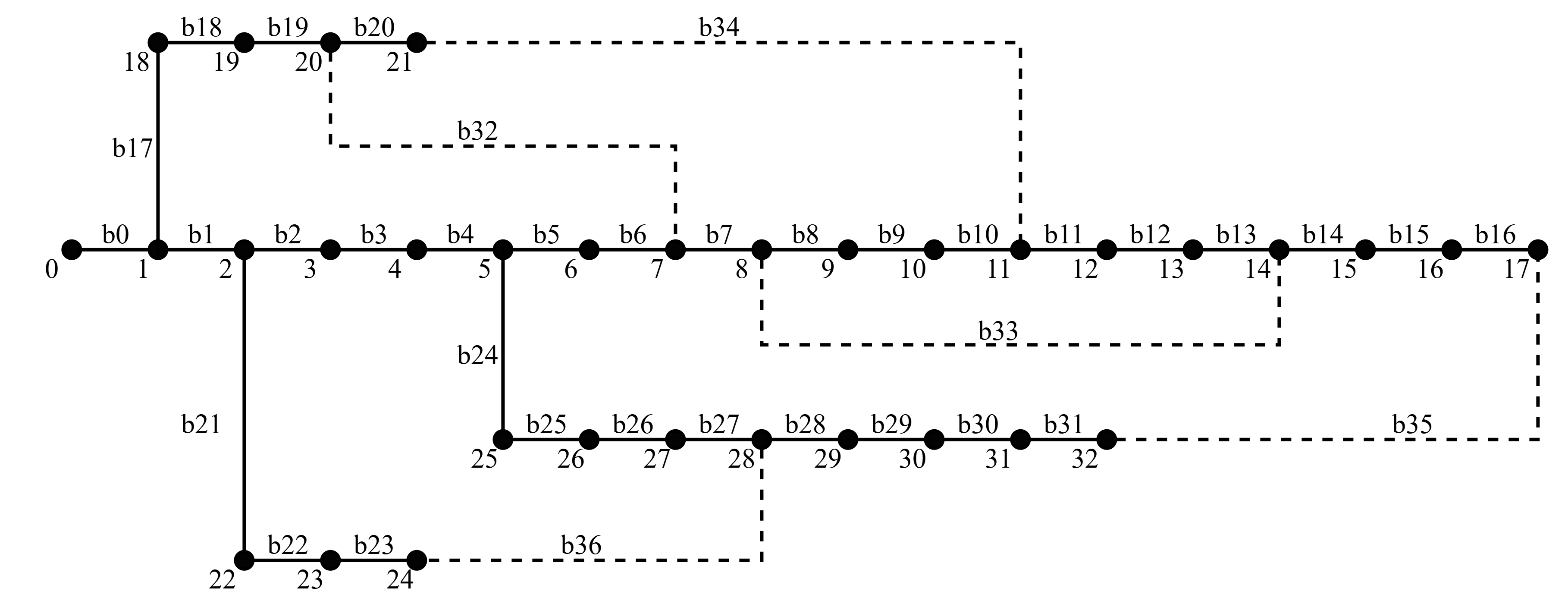}
    \caption{IEEE 33-bus distribution system in its initial radial configuration. Each line is modeled as switchable, with five switches initially open and the remaining switches closed.}
    \label{fig:IEEE33}
\end{figure*}

\subsection{Quantum Reinforcement Learning}
Quantum reinforcement learning (QRL) encompasses a broad class of approaches that incorporate quantum computation into the reinforcement learning pipeline~\cite{meyer2024survey}. These approaches range from quantum-inspired algorithms, which augment classical reinforcement learning with quantum-inspired subroutines~\cite{4579244}, to fully quantum algorithms in which both the state representation and learning procedure are implemented using quantum operations~\cite{PhysRevLett.117.130501}. For near-term gate-based quantum hardware, parameterized quantum circuit (PQC) based QRL methods have emerged as a promising approach for implementing trainable function approximators. In these methods, PQCs replace or augment classical neural networks, with the circuit parameters optimized through classical reinforcement learning procedures.

In this work, we begin with a classical neural network for approximating the action-value function $Q(s,a)$ and replace its classical backbone with a PQC-based quantum backbone, while retaining a classical decision head for producing the final action-value estimates. This architecture therefore forms a \textit{hybrid quantum-classical Q-network}, in which the PQC serves as the trainable feature-extraction backbone and its measured outputs are processed by a classical decision head to estimate the action values. The resulting approximation can be expressed as

\begin{equation}
Q(s,a)
\approx
Q_{\mathrm{hybrid}}(s,a;\boldsymbol{\theta}_{q},\boldsymbol{\theta}_{c}),
\end{equation}
where $\boldsymbol{\theta}_{q}$ denotes the trainable parameters of the quantum circuit and $\boldsymbol{\theta}_{c}$ denotes the trainable parameters of the classical decision head. In this work, the integration of this hybrid quantum-classical Q-network within the Deep Q-Learning framework is referred to as \textit{Quantum Deep Q-Learning}.

\section{Proposed DRL Algorithm}\label{sec:drl_algo}
\subsection{Environment}
Distribution network reconfiguration (DNR) can be formulated as a sequential decision-making problem suitable for reinforcement learning. In this formulation, an agent is trained to iteratively modify the distribution-network topology through branch-exchange operations with the objective of identifying configurations that improve network performance. The distribution system and its corresponding power-flow simulation together define the RL environment, through which the agent interacts with the network and evaluates the consequences of each switching decision.

In this work, the IEEE 33 (Fig. \ref{fig:IEEE33}) and 69-bus distribution systems are used as the RL environments. Each line in the network is modeled as switchable and is therefore associated with a binary open or closed status. Each system begins from a predefined initial radial configuration containing $n_open$ open switches, with the remaining switches closed. At each decision step, the agent performs a branch-exchange operation by closing one of the open switches and subsequently opening another switch on the resulting cycle. This operation modifies the network topology while preserving the number of open switches and maintaining radial operation.

Following each branch-exchange operation, a power-flow analysis is performed on the resulting network configuration and a reward is calculated based on its performance. The resulting configuration then defines the environment at the subsequent decision step, and this process continues until a terminal condition is reached. If an action produces a network configuration for which a valid power-flow solution cannot be obtained, the episode is terminated and the environment is reset to its initial configuration.

\begin{figure*}[h]
    \centering
\includegraphics[width=0.85\linewidth]{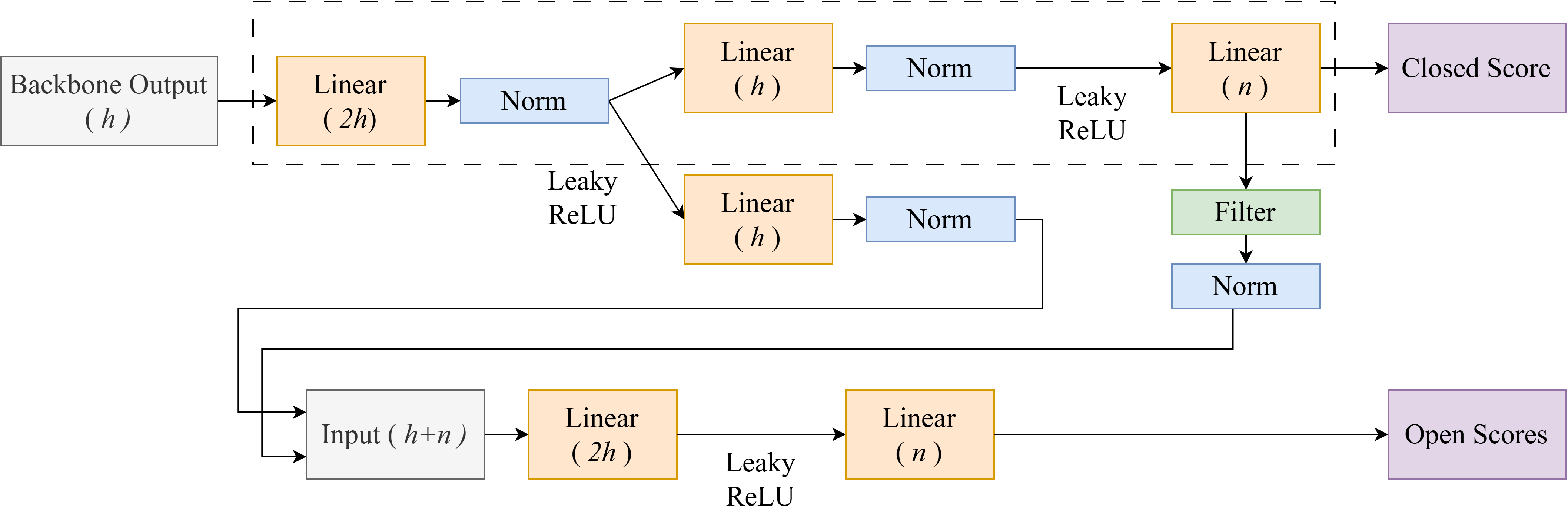}
    \caption{Double decision-head architecture used to implement the branch-exchange operation. The shared backbone representation is first passed to the closing head to score candidate closing actions. The closing-head output is then combined with the shared representation and passed to the opening head, allowing the opening decision to depend on the preceding closing decision.}
    \label{fig:ddh}
\end{figure*}

\subsection{State Representation}
At each decision step $t$, the state is constructed from the current status of the network switches and select power flow measurements. Specifically, the agent receives the voltage magnitude at each bus, the current flowing through each line, and the binary state of each switch. The resulting state is defined as

\begin{equation}
s_t = \left(s_t^{v},s_t^{i},s_t^{s}\right),
\end{equation}
where $s_t^{v}$ denotes the bus-voltage component,

\begin{equation}
s_t^{v} =
\left
(v_t^1,\ldots,v_t^\mathcal{B})
\mid
v_t^i \in \mathbb{R},
\right.
\end{equation}
with $v_t^i$ representing the voltage magnitude at bus $i$ of total buses $\mathcal{B}$, expressed in per-unit (p.u.). The line-current component is given by

\begin{equation}
s_t^{i} =
\left
(i_t^1,\ldots,i_t^\mathcal{L})
\mid
i_t^j \in \mathbb{R},
\right.
\end{equation}
where $i_t^j$ represents the current flowing through line $j$ of total lines $\mathcal{L}$, expressed in kA. Finally, the switch-status component is defined as

\begin{equation}
s_t^{s} \in \{0,1\}^{n},
\end{equation}
where $n$ is the number of switches and each element indicates whether the corresponding switch is open or closed. These features are provided to the neural network backbone for value-function approximation. More power flow measurements can be included in the state representation, but only these measurements were needed to reach the optimal configuration on the IEEE33 bus case.

\subsection{Action Space and Branch Exchange} \label{sec:action_space}
To execute the branch exchange operation the agent must make two sequential decisions: first, an open switch is selected and closed, and subsequently, a closed switch belonging to the resulting cycle is selected and opened. Let the branch-exchange action be represented by the tuple

\begin{equation}
a_t = (i,j),
\end{equation}
where $i$ denotes the switch selected for closing and $j$ denotes the switch selected for opening. At the first stage of the action, the Q-network produces a score for each switch corresponding to the closing decision. The scores are masked according to the current switch configuration so that only switches that are currently open can be selected. The $\epsilon$-greedy policy as defined in equation \ref{eq:epsilon_greedy} is then applied to the masked Q-values to select switch $i$.

Closing switch $i$ in a radial distribution network introduces a single cycle into the topology. The resulting cycle is identified from the modified network configuration and defines the set of valid switches for the second stage of the branch exchange. The Q-network's second decision head produces a score for each switch corresponding to the opening decision. These scores are again masked so that only switches belonging to the newly formed cycle can be selected. The $\epsilon$-greedy policy is then applied to the masked scores to select switch $j$ for opening. 

The action mask is applied to invalid actions by assigning a sufficiently large negative value prior to the $\arg\max$ operation (-1E5), preventing them from being selected during greedy action selection. During exploration, random actions are sampled only from the corresponding set of valid switches. The complete branch-exchange operation is therefore given by

\begin{equation}
s_t
\xrightarrow{\text{close }i}
\tilde{s}_t
\xrightarrow{\text{open }j}
s_{t+1},
\end{equation}
where $\tilde{s}_t$ denotes the intermediate network configuration containing the newly formed cycle and $s_{t+1}$ denotes the resulting radial configuration.

Following completion of the branch exchange, the resulting switch configuration is evaluated by the environment. The power flow is solved for the new configuration, and the resulting bus voltages and line currents form the corresponding electrical component of the next state. Thus, the transition can be expressed as

\begin{equation}
s_{t+1}=f(s_t,a_t),
\end{equation}
where $f$ applies the sequential branch-exchange operation and evaluates the resulting network through a power flow calculation.

\begin{figure*}[h]
    \centering
\includegraphics[width=0.85\linewidth]{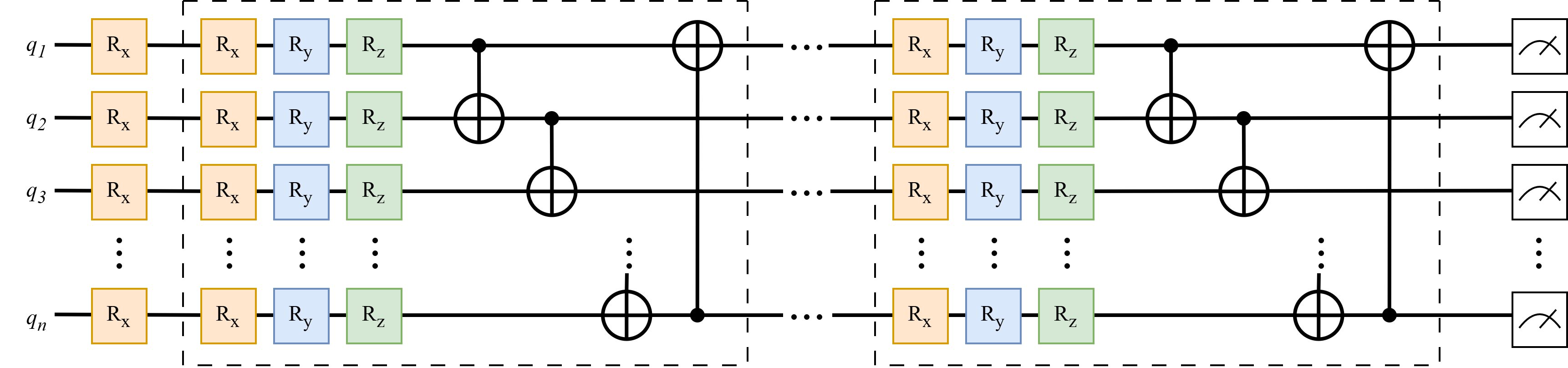}
    \caption{Parameterized quantum circuit (PQC) architecture used for each quantum layer of the quantum MLP. Each PQC consists of a data-embedding stage followed by a variational ansatz with four layers in all experiments. The complete quantum backbone consists of four sequential PQC layers of this form.}
    \label{fig:qmlp}
\end{figure*}

\subsection{Reward Function}
The reward function is designed to encourage the agent to discover network configurations that improve electrical performance. The total reward $R(s_t,a_t)\in\mathbb{R}$ is defined as

\begin{equation}
\label{rewardfunc}
R(s_t,a_t) =
\begin{cases}
0, & \text{if invalid},\\
R_e(s_t,a_t) + R_t(s_t,a_t), & \text{if valid}.
\end{cases}
\end{equation}
where $R_e$ denotes the electrical reward and $R_t$ denotes the topological reward. The electrical reward evaluates the power-flow characteristics of the resulting network configuration, while the topological reward provides an additional incentive for exploring configurations that differ from the initial network topology.

The electrical $R_e$ is defined as

\begin{equation}
R_e(s_t,a_t) = V_{\mathrm{mag}} + L_{\mathrm{loss}},
\end{equation}
where $V_{\mathrm{mag}}$ is the voltage-magnitude reward and $L_{\mathrm{loss}}$ is the active power-loss reward.

The voltage-magnitude reward is defined as

\begin{equation}
p_1\exp\left(
1-p_2\sum_{b=1}^{B}(1-v_t^b)^2
\right),
\end{equation}
where $v_t^b$ denotes the voltage magnitude at bus $b$ in per-unit. The reward is maximized when all bus voltages are equal to the nominal value of $1$ p.u. and decreases as the voltage magnitudes deviate from this value. The coefficient $p_2$ controls the sensitivity of the reward to voltage deviations, while the scaling factor $p_1$ scales the contribution of the voltage term to the total reward.

The active power-loss reward $L_{\mathrm{loss}}$ is defined as:

\begin{equation}
p_4\exp\left(
\frac{P_{\mathrm{init}}-P_t}{P_{\mathrm{init}}}
\cdot p_3
\right),
\end{equation}
where $P_{\mathrm{init}}$ is the total active power loss of the initial configuration and $P_t$ is the total active power loss of the resulting configuration. A reduction in active power loss relative to the initial configuration therefore increases the reward, while an increase in losses decreases it. The factor $p_3$ controls the sensitivity of the reward to relative changes in power loss and $p_4$ controls the overall sensitivity of the reward to line loss changes from the initial configuration. Together, the two power flow loss terms encourage the agent to identify configurations that simultaneously maintain bus voltages near their nominal values and reduce active power losses.

The topological reward  $R_t$ is calculated by 
\begin{equation}
R_t(s_t,a_t)
=
p_5\,d_H\left(s_t^s,s_0^s\right),
\label{eq:topological_reward}
\end{equation}
where $d_H(\cdot,\cdot)$ denotes the Hamming distance between the current switch configuration $s_t^s$ and the initial switch configuration $s_0^s$. The coefficient $p_5$ controls the contribution of the topological exploration reward to the total reward. This term encourages the agent to explore configurations that differ from the initial network topology.

\begin{figure}[h]
    \centering
\includegraphics[width=0.85\linewidth]{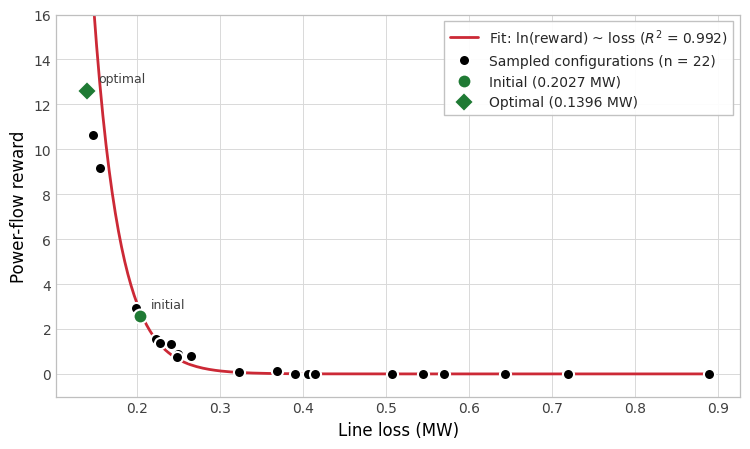}
    \caption{Power-flow reward (voltage and loss terms, no exploration bonus) against line loss for 22 radial IEEE 33-bus configurations. Reward parameters for this curve are listed in Table \ref{tab:mlp_33_params} of the Appendix. } 
    \label{fig:reward_curve}
\end{figure}

\subsection{Training Procedure}\label{sec:training_procedure}
Given the state representation, action space, and reward function described above, the agent is trained using a Double Deep Q-Network (Double DQN) as described in Section~\ref{sec:background}. During this training procedure each transition $(s_t,a_t,s_{t+1},r_t)$ is stored in a replay buffer. To improve the efficiency of experience replay, prioritized experience replay (PER)~\cite{schaul2016per} is employed rather than uniform sampling. Transitions with larger temporal-difference errors are assigned higher sampling probabilities and are therefore more likely to be selected for subsequent network updates. This allows transitions that provide a stronger learning signal to be revisited more frequently while retaining the benefits of replay for reducing temporal correlations between consecutive network configurations. Mini-batches of size $N_{\text{batch}}$ are sampled from the replay buffer and used to update the online Q-network.

\section{Q-Network Architecture}\label{sec:q_networks}
\subsection{Backbones}
The classical baseline uses a four-layer multilayer perceptron (MLP) backbone. The input is first projected to a 128-dimensional hidden representation, followed by three additional fully connected layers of width 128. LeakyReLU activations are applied between successive layers, producing a final 128-dimensional representation that is passed to the Q-network decision heads.

The quantum backbone replaces the fully connected hidden layers of the classical MLP with four sequential parameterized quantum circuit (PQC) layers, as illustrated in Fig.~\ref{fig:qmlp}. Each quantum layer follows the same general structure: its classical input features are transformed and encoded into a quantum state, a trainable variational ansatz is applied, and the resulting circuit is measured to produce a classical output representation. For the first three quantum layers, these measured outputs are passed to the subsequent quantum layer and re-encoded into a new quantum state. The output of the fourth and final quantum layer is instead passed to the classical portion of the Q-network.

Unlike the classical backbone, no additional nonlinear activation function is applied between successive quantum layers. Nonlinearity is instead introduced through the nonlinear transformation applied during feature embedding and through the quantum circuit and measurement process. Following the fourth quantum layer, a LeakyReLU activation is applied, after which a classical linear projection maps the measured quantum representation to the dimensionality required by the decision heads. This projection provides flexibility in the interface between the quantum backbone and the decision heads without requiring modifications to the quantum circuit architecture.

The number of qubits controls the width of each quantum layer and is treated as an architectural hyperparameter governing the capacity of the quantum backbone. All quantum MLP models are implemented using TorchQuantum and trained on NVIDIA H200 GPUs.

\begin{table*}[h]
\centering
\caption{MLP Models: 33 Bus Performance Statistics (4 seeds per model). Final reward is the seed mean $\pm$ sample standard deviation at the last training episode. Bold marks the smallest classical and quantum models to reach a hit rate of 1.0.}
\label{tab:33_mlp_stats}
\resizebox{\textwidth}{!}{%
\begin{tabular}{@{}lrrrrrrrrr@{}}
\toprule
\textbf{Metric} & \textbf{Classical (h=16)} & \textbf{Classical (h=32)} & \textbf{Classical (h=64)} & \textbf{Classical (h=128)} & \textbf{4-Qubit} & \textbf{8-Qubit} & \textbf{12-Qubit} & \textbf{16-Qubit} & \textbf{Unit} \\
\midrule
Total Parameters & 7,924 & \textbf{21,332} & 65,044 & 220,052 & 1,824 & 3,884 & \textbf{6,328} & 9,156 & count \\
Final Reward & 13.16 $\pm$ 0.74 & \textbf{13.12 $\pm$ 0.75} & 13.52 $\pm$ 0.40 & 13.37 $\pm$ 0.66 & 12.59 $\pm$ 0.50 & 12.98 $\pm$ 0.20 & \textbf{13.26 $\pm$ 0.39} & 13.37 $\pm$ 0.16 & ---- \\
\midrule
Minimum Loss & 0.1396 & \textbf{0.1396} & 0.1396 & 0.1396 & 0.1422 & 0.1422 & \textbf{0.1396} & 0.1396 & MW \\
Maximum Loss & 0.1422 & \textbf{0.1396} & 0.1396 & 0.1396 & 0.1422 & 0.1422 & \textbf{0.1396} & 0.1403 & MW \\
Mean Loss & 0.1403 & \textbf{0.1396} & 0.1396 & 0.1396 & 0.1422 & 0.1422 & \textbf{0.1396} & 0.1398 & MW \\
Median Loss & 0.1396 & \textbf{0.1396} & 0.1396 & 0.1396 & 0.1422 & 0.1422 & \textbf{0.1396} & 0.1396 & MW \\
Spread (Max-Min) & 0.0026 & \textbf{0.0} & 0.0 & 0.0 & 0.0 & 0.0 & \textbf{0.0} & 0.0007 & MW \\
Hit Rate & 0.75 & \textbf{1.0} & 1.0 & 1.0 & 0.0 & 0.0 & \textbf{1.0} & 0.75 & fraction \\
Median Gap & 0.0 & \textbf{0.0} & 0.0 & 0.0 & 1.8287 & 1.8287 & \textbf{0.0} & 0.0 & \% \\
Modal Agreement & 0.75 & \textbf{1.0} & 1.0 & 1.0 & 1.0 & 1.0 & \textbf{1.0} & 0.75 & fraction \\
\bottomrule
\end{tabular}%
}
\end{table*}

\subsubsection{Embedding Method}
Before entering the quantum backbone, the environment state $s_t$ is first mapped to an $n_q$-dimensional representation through a classical linear transformation, where $n_q$ denotes the number of qubits. This projection is necessary because the dimensionality of the environment state generally differs from the number of qubits used by the quantum backbone. The projected features are additionally multiplied by trainable scaling parameters before application of an arctangent nonlinearity, following the approach proposed in~\cite{pmlr-v148-lockwood21a}. The input feature associated with the $i$th qubit is therefore

\begin{equation}
x_i' = \arctan(x_i w_i),
\label{eq}
\end{equation}

\noindent where $x_i$ denotes the $i$th component of the projected classical representation and $w_i$ is its corresponding trainable weight. The transformed features are then used as rotation angles for single-qubit $R_X$ gates. The resulting encoded state is

\begin{equation}
|\psi(x')\rangle=
\bigotimes_{i=1}^{n_q}
R_X(x_i')|0\rangle,
\label{eq}
\end{equation}

This angle-encoding scheme provides a one-to-one mapping between the transformed classical features and the qubits of the circuit. It was selected for its simplicity and low circuit cost, requiring only a single $R_X$ gate per qubit and no additional entangling operations during the encoding stage. This limits circuit depth and the number of opportunities for hardware error, making the encoding well suited for subsequent evaluation on quantum hardware.

\subsubsection{Ansatz and Measurement}
Following angle encoding, each quantum MLP applies a trainable variational ansatz to the encoded state $|\psi(x')\rangle$. Each ansatz layer consists of three trainable single-qubit rotations applied to every qubit, followed by a circular CNOT entangling layer. Denoting the unitary corresponding to the $\ell$th ansatz layer by $U_{\mathrm{ansatz}}^{(\ell)}(\boldsymbol{\theta}^{(\ell)})$, the full four-layer variational transformation is

\begin{equation}
|\psi(x',\boldsymbol{\theta})\rangle
=
\left[
\prod_{\ell=1}^{4}
U_{\mathrm{ansatz}}^{(\ell)}
\left(\boldsymbol{\theta}^{(\ell)}\right)
\right]
|\psi(x')\rangle,
\label{eq:variational_ansatz}
\end{equation}
where $\boldsymbol{\theta}^{(\ell)}$ contains the trainable rotation parameters of the $\ell$th ansatz layer and $\boldsymbol{\theta}=\{\boldsymbol{\theta}^{(1)},\ldots,\boldsymbol{\theta}^{(4)}\}$ denotes the complete set of variational parameters.

Following the four-layer variational transformation, each qubit is measured through the expectation value of the Pauli-$Z$ observable. For the $i$th qubit, the resulting output is

\begin{equation}
z_i
=
\langle \psi(x',\boldsymbol{\theta})|
Z_i
|\psi(x',\boldsymbol{\theta})\rangle,
\qquad i=1,\ldots,n,
\label{eq:z_measurement}
\end{equation}
where $Z_i$ denotes the Pauli-$Z$ operator acting on qubit $i$. Measuring all $n_q$ qubits therefore produces the $n_q$-dimensional output representation

\begin{equation}
\mathbf{z}
=
[z_1,z_2,\ldots,z_n]^{\mathrm{T}}
\in \mathbb{R}^{n_q}.
\label{eq:quantum_representation}
\end{equation}

As previously stated, the quantum MLP consists of four sequential layers of this structure. For the first three layers, the measured representation $\mathbf{z}$ is passed forward as the input features to the subsequent quantum layer, where it is re-encoded into a new quantum state before application of the next variational ansatz. After the fourth quantum layer, the measured representation is passed to the classical decision head used to produce the Q-network outputs.

\subsection{Double Decision Head}

The output of the shared backbone is passed directly to a double decision head  (Fig \ref{fig:ddh}) that implements the branch-exchange operation. The decision head consists of separate closing and opening heads, with the closing decision evaluated first and its output subsequently incorporated into the opening decision. This structure allows the Q-network to model the dependency between the branch selected for closing and the branch subsequently selected for opening.

The shared backbone produces a hidden representation of dimension $h$, where $h$ is determined by the backbone architecture. The closing head first expands the shared representation from $h$ to $2h$ dimensions and applies layer normalization and a LeakyReLU activation. This representation is then passed through a linear layer to produce a score for each possible closing action. The resulting closing-action scores are subsequently combined with the shared representation to provide the input to the opening head. The opening head processes this combined representation and produces a score for each possible opening action. 

\begin{figure*}[h]
    \centering
    \includegraphics[width=1\linewidth]{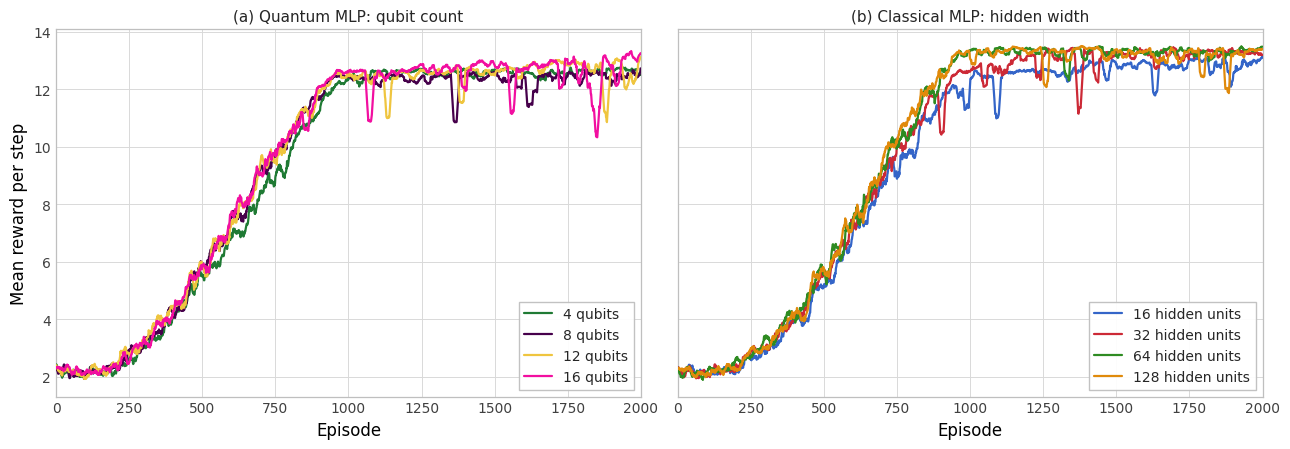}
    \caption{Reward curves for the classical and quantum ablation studies: (a) Quantum MLP with 4, 8, 12 and 16 qubits; (b) Classical MLP with 16, 32, 64 and 128 hidden units. Each curve is the mean over 4 seeds, smoothed with a 20-episode moving average. Both panels share the same x-axis (0–2,000 episodes) and y-axis, so the two families can be compared directly.}
    \label{fig:ablation_reward}
\end{figure*}

\begin{figure*}[h]
    \centering
    \includegraphics[width=1\linewidth]{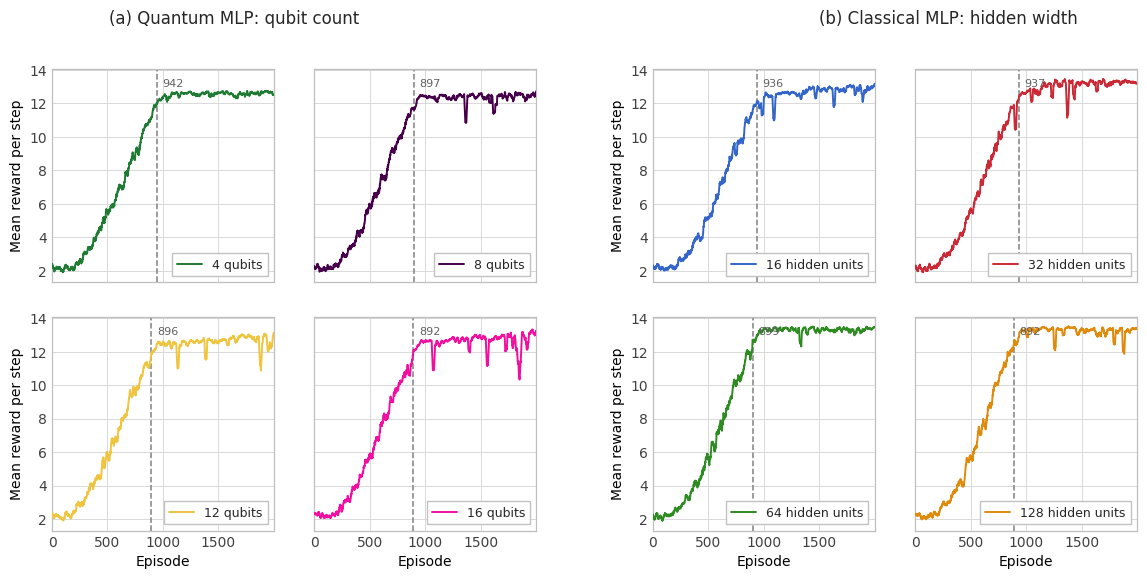}
    \caption {The ablation study reward curves with one panel per configuration: (a) Quantum MLP by qubit count, (b) Classical MLP by hidden width. The dashed vertical line and its label mark the first episode where the smoothed seed-mean reward reaches 95\% of that configuration's plateau, defined as the mean reward over the final 50\% of episodes.}
    \label{fig:ablation_reward_split}
\end{figure*}

\section{Results} \label{sec:results}
The proposed Q-network architectures are evaluated on the IEEE 33 and 69 bus systems. For the IEEE 33 bus system, ablation studies are conducted to examine the effects of model capacity on performance. For the quantum models, the circuit width is varied from 4 to 16 qubits in increments of four, while for the classical models, the hidden dimension is varied from 16 to 128. A hidden dimension of 16 is selected as the initial classical baseline because its parameter count is comparable to those of the larger quantum models, enabling comparisons between architectures with similar model capacities. Due to the cost of quantum circuit simulation, the full model-size ablation was not conducted on the 69 bus system. 

For the IEEE 33-bus system, multiple random seeds are used to assess training stability and variation in solution quality. The IEEE 69-bus system is evaluated using a single seed with a 12-qubit model, corresponding to the qubit count that exhibited the most stable performance across seeds in the IEEE 33-bus ablation study. Within each bus system, the classical and quantum models are trained under identical training conditions, with grid-specific parameter settings provided in the Appendix. Following simulation, selected trained quantum models are executed on IonQ Forte Enterprise-1 hardware, and their inference outputs are compared with the corresponding noiseless-simulation results to characterize the effects of quantum hardware noise on model inference.

\subsection{IEEE 33 Bus Case}
\subsubsection{Final Performance and Parameter Efficiency}
Table~\ref{tab:33_mlp_stats} summarizes the performance of the classical and hybrid quantum Q-network models across the four random seeds. For each seed, the final reward is calculated as the mean reward obtained during the final episode. The reported final reward is then computed as the mean of these four seed-level values, with the accompanying standard deviation representing variation across seeds. Thus, final reward measures the average quality of the solutions produced at the end of training. The metrics are computed using the SciPy package in Python \cite{2020SciPy-NMeth}. 

Under this definition, the quantum models exhibited an increasing mean final reward with qubit count, rising from $12.59 \pm 0.50$ for the 4 qubit model to $13.37 \pm 0.16$ for the 16 qubit model. The decrease in standard deviation from $0.50$ to $0.16$ also indicates reduced variation in final reward across the evaluated seeds. However, the increase in final reward was not accompanied by a monotonic increase in exact-optimum attainment. The 12 qubit model identified the true optimum in all four seeds (hit rate $=1.0$), whereas the 16 qubit model did so in three of four seeds (hit rate $=0.75$). Thus, within the evaluated quantum widths, increasing model size improved the average final reward but did not produce a corresponding improvement in the frequency of exact-optimum solutions. Additional seeds would be required to determine whether this difference is systematic.

The quantum models also exhibited a parameter efficiency advantage relative to the classical models. Among the classical configurations, the hidden dimension 32 model was the smallest model to identify the true optimum across all four seeds. The 12 qubit model achieved the same perfect hit rate with 6,328 trainable parameters, compared with 21,332 parameters for the classical hidden dimension 32 model, corresponding to a \textbf{70.3\%} reduction in parameter count. The difference is also reflected in final reward: the 12  and 16 qubit models achieved mean final rewards of $13.26 \pm 0.39$ and $13.37 \pm 0.16$, respectively, compared with $13.12 \pm 0.75$ for the classical hidden-dimension-32 model. These results show that the quantum models can achieve equal or greater performance with fewer trainable parameters.

Unlike the quantum models, the classical models did not exhibit the same monotonic relationship between model size and final reward. Mean final reward increased with hidden dimension up to 64, after which performance decreased for the 128 hidden dimension model. Notably, both the 128 hidden dimension model and the 16 qubit quantum model achieved a mean final reward of $13.37$, despite using 220,052 and 9,156 trainable parameters respectively. The classical model therefore used approximately \textbf{2,303\% more parameters} to achieve the same mean final reward and also exhibited greater variation across seeds. These observations motivate extending the quantum width ablation to larger circuits to determine whether the increasing reward trend observed over 4-16 qubits eventually saturates or reverses. Such experiments could additionally investigate whether optimization effects, including gradient degradation or barren-plateau behavior, contribute to diminishing returns at larger circuit widths.

\subsubsection{Training Dynamics}
Beyond final performance, we examined whether model size affected training speed. Figure~\ref{fig:ablation_reward} shows side-by-side reward curves for the different qubit counts and classical hidden dimensions, while Figure~\ref{fig:ablation_reward_split} separates the individual curves into a grid. The vertical dashed line in each pane of Figure \ref{fig:ablation_reward_split} indicates the convergence episode. Convergence is defined as the first episode at which the smoothed seed-mean reward reaches 95\% of the configuration's plateau reward, where the plateau is defined as the mean reward over the final 50\% of training episodes.

Although the 12 and 16 qubit models achieved higher mean final episode rewards, the classical models maintained slightly higher average rewards over the final 50\% of training episodes. The quantum models nevertheless exhibited a steeper increase in reward near the end of training. More importantly, these differences in reward trajectories did not translate into substantial differences in convergence speed. The quantum models converged between episodes 892 and 942, while the classical models converged between episodes 892 and 937. Thus, across the evaluated architectures, increasing model capacity did not substantially accelerate convergence. The 16 qubit quantum model and the 128 hidden dimension classical model both converged at episode 892, despite containing only 9,156 and 220,052 trainable parameters, respectively. The quantum model therefore reached the same convergence criterion with approximately \textbf{95.8\% fewer parameters}.

Taken together, the IEEE 33-bus results demonstrate that the evaluated quantum architectures achieved competitive final performance with substantially fewer trainable parameters than the classical architectures. Increasing the quantum width from 4 to 16 qubits was associated with higher mean final reward and lower variation across seeds, although exact-optimum attainment peaked at 12 qubits rather than continuing to increase at 16 qubits. The classical models exhibited a non-monotonic relationship between hidden dimension and final reward, while the largest classical model required substantially more parameters to achieve a final reward comparable to that of the 16-qubit model. Finally, the similar convergence ranges across architectures indicate that this parameter-efficiency advantage did not come at the cost of slower convergence in the present experiments.

\begin{figure*}[h]
    \centering
    \includegraphics[width=1.0\linewidth]{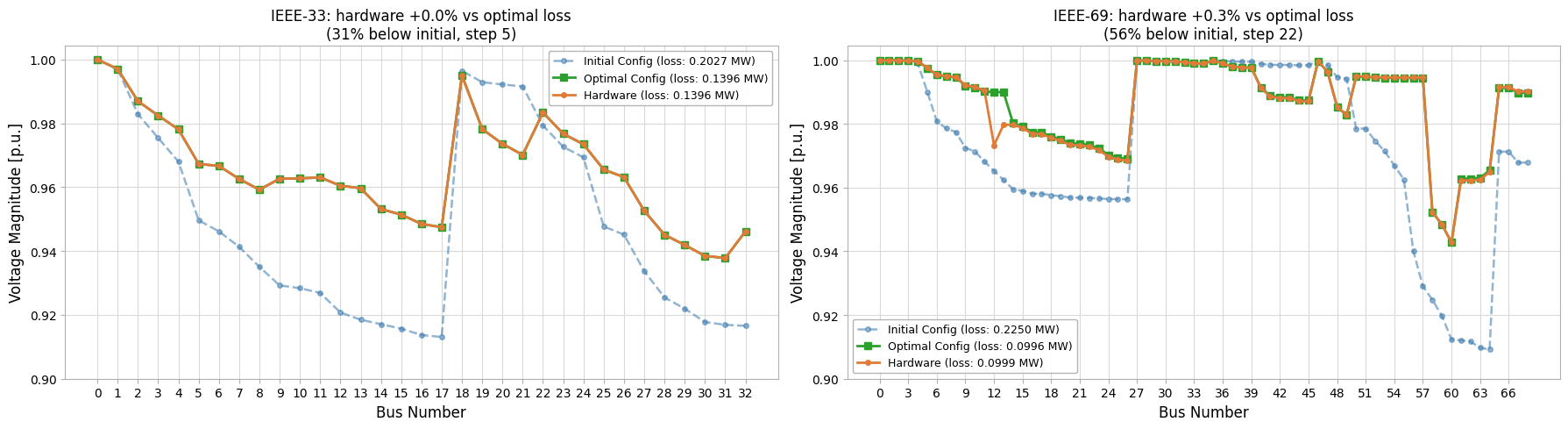}
    \caption{Comparison of bus voltage magnitude profiles for the IEEE-33 (left) and IEEE-69 (right) feeders. Each panel shows the initial radial configuration (dashed), the known optimal configuration, and the best configuration reached by the quantum MLP DQN policy executed on IonQ's Forte Enterprise 1 system. All four variational layers were evaluated on hardware using 12 qubits and 2000 shots per circuit, with the agent allowed a maximum of 24 steps. For the IEEE-33 system, the hardware-executed policy recovers the true optimal configuration, while for the IEEE-69 system, the best hardware solution achieves a total line loss within 0.3\% of the known optimum.}
    \label{fig:profilecomparison}
\end{figure*}

\subsection{Hardware Evaluation}

Figure~\ref{fig:profilecomparison} compares the initial voltage profiles with those obtained after reconfiguration using the trained 12 qubit quantum models under ideal simulation and quantum hardware execution for the IEEE 33 and 69 bus systems. For both systems, hardware inference was performed using 2,000 measurement shots on IonQ Forte Enterprise-1. Under ideal simulation, the quantum models identify the known optimal configurations for both systems, corresponding to line losses of $0.1396$~MW and $0.0996$~MW for the IEEE 33 and 69 bus systems, respectively. When executed on quantum hardware, the IEEE 33 bus model again identifies the known optimal configuration, achieving a line loss of $0.1396$~MW, while the IEEE 69 bus model produces a configuration with a line loss of $0.0999$~MW.

These results indicate that the trained quantum models retain their ability to produce high-quality reconfiguration decisions when transferred from ideal simulation to physical quantum hardware. For the IEEE 33-bus system, hardware execution reproduces the optimal simulated result despite finite-shot sampling and hardware noise. For the larger IEEE 69-bus system, hardware execution results in a modest increase in line loss of $0.0003$~MW, corresponding to approximately $0.30\%$ relative to the simulated optimum. The close agreement between the simulated and hardware-derived configurations provides evidence that the proposed hybrid quantum-classical Q-network can maintain its inference performance under the noise and sampling effects present on current quantum hardware.

\section{Conclusion} \label{sec:conclusion}

This work investigated the use of hybrid quantum--classical deep reinforcement learning for distribution network reconfiguration, with an emphasis on the performance and parameter efficiency of quantum Q-network architectures. A hybrid Double DQN framework was developed in which a parameterized quantum circuit replaces the classical MLP backbone while retaining a classical decision architecture for performing sequential branch-exchange operations. The proposed framework was evaluated through an ablation study on the IEEE 33-bus distribution system, comparing quantum models of increasing circuit width with classical MLPs of increasing hidden dimension.

The results demonstrate that the quantum models can achieve competitive or improved solution quality with substantially fewer trainable parameters than the evaluated classical architectures. Increasing the quantum model from 4 to 16 qubits resulted in an increase in mean final reward from $12.59 \pm 0.50$ to $13.37 \pm 0.16$, while also reducing variation across the evaluated seeds. However, exact-optimum attainment did not increase monotonically with circuit width. The 12-qubit model identified the true optimal configuration across all four seeds, while the 16-qubit model achieved a hit rate of $0.75$, indicating that increases in model capacity and mean final reward do not necessarily correspond to more frequent recovery of the exact optimum.

The strongest distinction between the quantum and classical models was observed in parameter efficiency. The 12-qubit model achieved a perfect hit rate using 6,328 trainable parameters, compared with 21,332 parameters for the smallest classical model to achieve the same hit rate, representing a $70.3\%$ reduction in trainable parameters. Similarly, the 16-qubit model achieved the same mean final reward as the 128-hidden-dimension classical model while using 9,156 parameters compared with 220,052. Despite this substantial difference in model size, the evaluated quantum and classical architectures exhibited similar convergence ranges. These results indicate that, for the IEEE 33-bus case considered in this work, the reduction in trainable parameters achieved by the quantum architecture did not come at the cost of slower convergence or reduced final solution quality.

Execution on trapped ion quantum hardware provides an initial assessment of how the trained quantum policy behaves beyond ideal simulation. For the IEEE 33 bus system, the 12 qubit model identifies the known optimal configuration both in simulation and when executed on IonQ Forte Enterprise-1 hardware. For the IEEE 69 bus system, hardware execution results in a line loss approximately $0.30\%$ above the simulated optimum. Although the optimal configuration is not recovered for the larger system, the small degradation in solution quality demonstrates that the trained quantum policy can retain strong performance under the noise and finite-shot sampling present on IonQ hardware.

Future work should extend this analysis to bus systems whose radial configuration spaces are too large to be exhaustively enumerated using classical methods. Existing benchmarks span configuration spaces ranging from approximately $5\times10^4$ states for the IEEE 33-bus system to more than $10^{35}$ states in the Ababei benchmark set~\cite{ababei2011reconfiguration,ababei2009reds}. Evaluating systems across this range would help establish the configuration-space sizes for which branch-exchange DQN remains effective and, importantly, identify the regimes in which classical and quantum approaches begin to degrade or fail. In this context, the rate of performance degradation as the configuration space grows may provide a more meaningful measure of DQN scalability than bus count alone, since the underlying learning problem is governed more directly by the number of feasible network configurations than by the number of buses.

Additionally, the reward used includes a power loss term, a voltage term, and an exploration bonus based on the Hamming weight distance to initial configuration. Since the exploration bonus is not potential shaped, it is not guaranteed to preserve the optimal policy \cite{ng1999policy}. It is important to study the performance of the models with various magnitudes of this metric - since this should have a nontrivial scaling with the bus size. Further work should verify this directly, checking that the optimum is the reward maximum over a range of exploration bonus magnitudes. 

Further, we would like to scale these simulations in terms of seeds and qubit counts.
Here the comparisons rest on 4 seeds per model, which can expose large effects, but not necessarily statistical differences between close values, as studies have found some sensitivity to seeds in DRL \cite{henderson2018matters}. Extending beyond 16 qubits is crucial both to understand performance saturation as well as potential training degradation caused by barren plateaus \cite{mcclean2018barren}, as this informs circuit and model design. In terms of the hardware execution, the 0.3\% degradation observed on QPU for the 69 bus case motivates noise aware training \cite{wang2022quantumnat,skolik2023robustness}, where noise is injected during training such that the model prefers relative ordering of expectation values over absolute magnitude. Then, the implementation and shots-tradeoff of applying error mitigation techniques should be considered for running DRL models on quantum hardware \cite{temme2017zne,cai2023qem}.

Overall, these results demonstrate that hybrid quantum--classical reinforcement learning can provide a parameter-efficient approach to distribution network reconfiguration while maintaining competitive solution quality. The combination of multi-seed simulation and execution on physical quantum hardware provides evidence for the potential of hybrid quantum Q-networks for power-system optimization, while also identifying scalability and hardware robustness as important directions for continued investigation.

\section*{Acknowledgment}\label{sec:acknowledgement}
We thank Shaked Regev and Phil Lotshaw of ORNL and Elias Kokkas of IonQ for their insights throughout this study. This article was prepared by EPB Quantum using Federal funds under award GONANB24D219-1 "Empowering the Future Grid: Quantum Training for Software Innovation \& Utility Applications." from the NIST Grant, U.S. Department of Commerce. The statement, findings, conclusions, and recommendations are those of the authors.

\bibliographystyle{IEEEtran}
\bibliography{bibliography}

@article{Schuetz_2022,
   title={Combinatorial optimization with physics-inspired graph neural networks},
   volume={4},
   ISSN={2522-5839},
   url={http://dx.doi.org/10.1038/s42256-022-00468-6},
   DOI={10.1038/s42256-022-00468-6},
   number={4},
   journal={Nature Machine Intelligence},
   publisher={Springer Science and Business Media LLC},
   author={Schuetz, Martin J. A. and Brubaker, J. Kyle and Katzgraber, Helmut G.},
   year={2022},
   month=apr, pages={367–377} }

@article{lotfi2024dnr,
  author={H. Lotfi and M. E. Hajiabadi and H. Parsadust},
  title={{Power Distribution Network Reconfiguration Techniques: A Thorough Review}},
  journal={Sustainability},
  volume={16},
  number={23},
  pages={10307},
  month={Nov.},
  year={2024},
  note={[Online]. Available: \url{https://doi.org/10.3390/su162310307}}
}

@article{skolfield2020operations,
  author = {Skolfield, J. Kyle and Escobedo, Adolfo R.},
  title = {Operations research in optimal power flow: A guide to recent and emerging methodologies and applications},
  journal = {European Journal of Operational Research},
  volume = {286},
  number = {2},
  pages = {379--395},
  year = {2020},
  doi = {10.1016/j.ejor.2020.03.031}
}

@article{oh2020online,
title = {Online reconfiguration scheme of self-sufficient distribution network based on a reinforcement learning approach},
journal = {Applied Energy},
volume = {280},
pages = {115900},
year = {2020},
issn = {0306-2619},
doi = {https://doi.org/10.1016/j.apenergy.2020.115900},
url = {https://www.sciencedirect.com/science/article/pii/S0306261920313672},
author = {Seok Hwa Oh and Yong Tae Yoon and Seung Wan Kim},
}

@article{Chen2020VQC,
  author = {Chen, Samuel Yen-Chi and Yang, Chao-Han Huck and Qi, Jun and Chen, Pin-Yu and Ma, Xiaoli and Goan, Hsi-Sheng},
  title = {Variational Quantum Circuits for Deep Reinforcement Learning},
  journal = {IEEE Access},
  year = {2020},
  volume = {8},
  pages = {141007--141024},
  doi = {10.1109/ACCESS.2020.3010470},
  issn = {2169-3536},
  month = {July}
}

@article{Gholizadeh2023,
  author = {Gholizadeh, Nastaran and Kazemi, Nazli and Musilek, Petr},
  title = {A {C}omparative {S}tudy of {R}einforcement {L}earning {A}lgorithms for {D}istribution {N}etwork {R}econfiguration {W}ith {D}eep {Q}-{L}earning-{B}ased {A}ction {S}ampling},
  journal = {IEEE Access},
  volume = {11},
  pages = {13714--13723},
  year = {2023},
  month = {February},
  doi = {10.1109/ACCESS.2023.3243549}
}

@article{Wang2021,
  author = {Wang, Beibei and Zhu, Hong and Xu, Honghua and Bao, Yuqing and Di, Huifang},
  title = {Distribution {N}etwork {R}econfiguration {B}ased on {N}oisyNet {D}eep {Q}-{L}earning {N}etwork},
  journal = {IEEE Access},
  volume = {9},
  pages = {90358--90365},
  year = {2021},
  month = {June},
  doi = {10.1109/ACCESS.2021.3089625}
}

@article{Gautam2023,
  author = {Gautam, Mukesh and Bhusal, Narayan and Benidris, Mohammed},
  title = {Deep {Q}-{L}earning-based {D}istribution {N}etwork {R}econfiguration for {R}eliability {I}mprovement},
  journal = {arXiv preprint arXiv:2305.01180},
  year = {2023},
  month = {May},
  note = {[eess.SY]}
}

@inproceedings{Di2020,
  author = {Di, Huifang and Bao, Yuqing and Wang, Beibei},
  title = {Research on {D}istribution {N}etwork {R}econfiguration {B}ased on {D}eep {Q}-learning {N}etwork},
  booktitle = {Proceedings of the 4th IEEE Conference on Energy Internet and Energy System Integration (EI2)},
  year = {2020},
  month = {October--November},
  address = {Wuhan, China},
  pages = {2728--2732},
  doi = {10.1109/EI250167.2020.9346939}
}

@article{Skolik2022,
  author = {Skolik, Andrea and Jerbi, Sofiene and Dunjko, Vedran},
  title = {Quantum agents in the {G}ym: a variational quantum algorithm for deep {Q}-learning},
  journal = {Quantum},
  volume = {6},
  pages = {720},
  year = {2022},
  month = {May},
  doi = {10.22331/q-2022-05-13}
}

@article{Lockwood2020,
  title={Reinforcement Learning with Quantum Variational Circuits},
  author={Lockwood, Owen and Si, Mei},
  journal={arXiv preprint arXiv:2008.07524v3},
  year={2020},
  month={August},
  archivePrefix={arXiv},
  primaryClass={quant-ph}
}

@article{meyer2024survey,
  title         = {A Survey on Quantum Reinforcement Learning},
  author        = {Meyer, Nico and Ufrecht, Christian and Periyasamy, Maniraman
                   and Scherer, Daniel D. and Plinge, Axel and Mutschler, Christopher},
  journal       = {arXiv preprint arXiv:2211.03464},
  year          = {2024},
  doi           = {10.48550/arXiv.2211.03464},
  eprint        = {2211.03464},
  archivePrefix = {arXiv},
  primaryClass  = {quant-ph},
  version       = {2}
}

@ARTICLE{4579244,
  author={Dong, Daoyi and Chen, Chunlin and Li, Hanxiong and Tarn, Tzyh-Jong},
  journal={IEEE Transactions on Systems, Man, and Cybernetics, Part B (Cybernetics)}, 
  title={Quantum Reinforcement Learning}, 
  year={2008},
  volume={38},
  number={5},
  pages={1207-1220},
  doi={10.1109/TSMCB.2008.925743}}

@article{PhysRevLett.117.130501,
  title = {Quantum-Enhanced Machine Learning},
  author = {Dunjko, Vedran and Taylor, Jacob M. and Briegel, Hans J.},
  journal = {Phys. Rev. Lett.},
  volume = {117},
  issue = {13},
  pages = {130501},
  numpages = {6},
  year = {2016},
  month = {Sep},
  publisher = {American Physical Society},
  doi = {10.1103/PhysRevLett.117.130501},
  url = {https://link.aps.org/doi/10.1103/PhysRevLett.117.130501}
}

@article{chen2024deep,
  author  = {Chen, H. Y. and Chang, Y. J. and Liao, S. W. and others},
  title   = {Deep Q-learning with hybrid quantum neural network on solving maze problems},
  journal = {Quantum Machine Intelligence},
  volume  = {6},
  pages   = {2},
  year    = {2024},
  month   = jan,
  doi     = {10.1007/s42484-023-00137-w}
}

@InProceedings{pmlr-v148-lockwood21a,
  title = 	 {Playing Atari with Hybrid Quantum-Classical Reinforcement Learning},
  author =       {Lockwood, Owen and Si, Mei},
  booktitle = 	 {NeurIPS 2020 Workshop on Pre-registration in Machine Learning},
  pages = 	 {285--301},
  year = 	 {2021},
  editor = 	 {Bertinetto, Luca and Henriques, João F. and Albanie, Samuel and Paganini, Michela and Varol, Gül},
  volume = 	 {148},
  series = 	 {Proceedings of Machine Learning Research},
  month = 	 {11 Dec},
  publisher =    {PMLR},
  url = 	 {https://proceedings.mlr.press/v148/lockwood21a.html}
}

@article{baranwu1989reconfiguration,
  author  = {Baran, Mesut E. and Wu, Felix F.},
  title   = {Network reconfiguration in distribution systems for loss
             reduction and load balancing},
  journal = {IEEE Transactions on Power Delivery},
  volume  = {4},
  number  = {2},
  pages   = {1401--1407},
  year    = {1989},
  doi     = {10.1109/61.25627}
}

@article{civanlar1988feeder,
  author  = {Civanlar, S. and Grainger, J. J. and Yin, H. and Lee, S. S. H.},
  title   = {Distribution feeder reconfiguration for loss reduction},
  journal = {IEEE Transactions on Power Delivery},
  volume  = {3},
  number  = {3},
  pages   = {1217--1223},
  year    = {1988},
  doi     = {10.1109/61.193906}
}

@article{lavorato2012radiality,
  author  = {Lavorato, Marina and Franco, John F. and Rider, Marcos J. and
             Romero, Ruben},
  title   = {Imposing Radiality Constraints in Distribution System
             Optimization Problems},
  journal = {IEEE Transactions on Power Systems},
  volume  = {27},
  number  = {1},
  pages   = {172--180},
  year    = {2012},
  doi     = {10.1109/TPWRS.2011.2161349}
}

@article{bienstock2019nphard,
  author  = {Bienstock, Daniel and Verma, Abhinav},
  title   = {Strong {NP}-hardness of {AC} power flows feasibility},
  journal = {Operations Research Letters},
  volume  = {47},
  number  = {6},
  pages   = {494--501},
  year    = {2019},
  doi     = {10.1016/j.orl.2019.08.009}
}

@article{morton2000bruteforce,
  author  = {Morton, A. B. and Mareels, I. M. Y.},
  title   = {An efficient brute-force solution to the network
             reconfiguration problem},
  journal = {IEEE Transactions on Power Delivery},
  volume  = {15},
  number  = {3},
  pages   = {996--1000},
  year    = {2000},
  doi     = {10.1109/61.871365}
}

@article{zhang2018dlreview,
  author  = {Zhang, Dongxia and Han, Xiaoqing and Deng, Chunyu},
  title   = {Review on the research and practice of deep learning and reinforcement learning in smart grids},
  journal = {CSEE Journal of Power and Energy Systems},
  volume  = {4},
  number  = {3},
  pages   = {362--370},
  year    = {2018},
  doi     = {10.17775/CSEEJPES.2018.00520}
}

@article{duan2020voltagecontrol,
  author  = {Duan, Jiajun and Shi, Di and Diao, Ruisheng and Li, Haifeng and
             Wang, Zhiwei and Zhang, Bei and Bian, Desong and Yi, Zhehan},
  title   = {Deep-Reinforcement-Learning-Based Autonomous Voltage Control
             for Power Grid Operations},
  journal = {IEEE Transactions on Power Systems},
  volume  = {35},
  number  = {1},
  pages   = {814--817},
  year    = {2020},
  doi     = {10.1109/TPWRS.2019.2941134}
}

@book{sutton2018rl,
  author    = {Sutton, Richard S. and Barto, Andrew G.},
  title     = {Reinforcement Learning: An Introduction},
  edition   = {2nd},
  publisher = {MIT Press},
  address   = {Cambridge, MA, USA},
  year      = {2018}
}

@article{watkins1992qlearning,
  author  = {Watkins, Christopher J. C. H. and Dayan, Peter},
  title   = {{Q}-learning},
  journal = {Machine Learning},
  volume  = {8},
  number  = {3--4},
  pages   = {279--292},
  year    = {1992},
  doi     = {10.1007/BF00992698}
}

@article{mnih2015dqn,
  author  = {Mnih, Volodymyr and Kavukcuoglu, Koray and Silver, David and
             Rusu, Andrei A. and Veness, Joel and Bellemare, Marc G. and
             Graves, Alex and Riedmiller, Martin and Fidjeland, Andreas K. and
             Ostrovski, Georg and Petersen, Stig and Beattie, Charles and
             Sadik, Amir and Antonoglou, Ioannis and King, Helen and
             Kumaran, Dharshan and Wierstra, Daan and Legg, Shane and
             Hassabis, Demis},
  title   = {Human-level control through deep reinforcement learning},
  journal = {Nature},
  volume  = {518},
  number  = {7540},
  pages   = {529--533},
  year    = {2015},
  doi     = {10.1038/nature14236}
}

@inproceedings{schaul2016per,
  author    = {Schaul, Tom and Quan, John and Antonoglou, Ioannis and
               Silver, David},
  title     = {Prioritized Experience Replay},
  booktitle = {Proc. 4th International Conference on Learning
               Representations (ICLR)},
  year      = {2016}
}

@inproceedings{fortunato2018noisynet,
  author    = {Fortunato, Meire and Azar, Mohammad Gheshlaghi and
               Piot, Bilal and Menick, Jacob and Osband, Ian and
               Graves, Alex and Mnih, Vlad and Munos, Remi and
               Hassabis, Demis and Pietquin, Olivier and
               Blundell, Charles and Legg, Shane},
  title     = {Noisy Networks for Exploration},
  booktitle = {Proc. 6th International Conference on Learning
               Representations (ICLR)},
  year      = {2018}
}

@article{mcclean2018barren,
  author  = {McClean, Jarrod R. and Boixo, Sergio and Smelyanskiy, Vadim N. and
             Babbush, Ryan and Neven, Hartmut},
  title   = {Barren plateaus in quantum neural network training landscapes},
  journal = {Nature Communications},
  volume  = {9},
  pages   = {4812},
  year    = {2018},
  doi     = {10.1038/s41467-018-07090-4}
}

@inproceedings{jerbi2021policies,
  author    = {Jerbi, Sofiene and Gyurik, Casper and Marshall, Simon C. and
               Briegel, Hans J. and Dunjko, Vedran},
  title     = {Parametrized Quantum Policies for Reinforcement Learning},
  booktitle = {Advances in Neural Information Processing Systems 34 (NeurIPS)},
  pages     = {28362--28375},
  year      = {2021}
}

@article{henderson2018matters,
    author = {Henderson, P. and Islam, R. and Bachman, P. and Pineau, J. and Precup, D. and Meger, D},
    title = {Deep Reinforcement Learning That Matters},
    journal = {Proceedings of the AAAI Conference on Artificial Intelligence},
    year = {2018},
    volume = {32},
    doi = {https://doi.org/10.1609/aaai.v32i1.11694}
}

@article{eskandarpour2020gridprimer,
  author  = {Eskandarpour, Rozhin and Ghosh, Kumar J. B. and Khodaei, Amin and
             Paaso, Aleksi and Zhang, Liuxi},
  title   = {Quantum-Enhanced Grid of the Future: A Primer},
  journal = {IEEE Access},
  volume  = {8},
  pages   = {188993--189002},
  year    = {2020},
  doi     = {10.1109/ACCESS.2020.3031595}
}

@article{golestan2023qcpower,
  author  = {Golestan, Saeed and Habibi, M. R. and Mousavi, S. M. Sajjad and
             Guerrero, Josep M. and Vasquez, Juan C.},
  title   = {Quantum computation in power systems: An overview of recent advances},
  journal = {Energy Reports},
  volume  = {9},
  pages   = {584--596},
  year    = {2023},
  doi     = {10.1016/j.egyr.2022.11.185}
}

@article{nara1992ga,
  author  = {Nara, Koichi and Shiose, Atsushi and Kitagawa, Minoru and
             Ishihara, Toshihisa},
  title   = {Implementation of genetic algorithm for distribution systems
             loss minimum re-configuration},
  journal = {IEEE Transactions on Power Systems},
  volume  = {7},
  number  = {3},
  pages   = {1044--1051},
  year    = {1992},
  doi     = {10.1109/59.207317}
}

@inproceedings{vanhasselt2016deep,
  title     = {Deep Reinforcement Learning with Double Q-Learning},
  author    = {van Hasselt, Hado and Guez, Arthur and Silver, David},
  booktitle = {Proceedings of the AAAI Conference on Artificial Intelligence},
  volume    = {30},
  number    = {1},
  pages     = {2094--2100},
  year      = {2016},
  doi       = {10.1609/aaai.v30i1.10295}
}

@article{temme2017zne,
  author  = {Temme, Kristan and Bravyi, Sergey and Gambetta, Jay M.},
  title   = {Error Mitigation for Short-Depth Quantum Circuits},
  journal = {Physical Review Letters},
  volume  = {119},
  number  = {18},
  pages   = {180509},
  year    = {2017}
}

@article{cai2023qem,
  author  = {Cai, Zhenyu and Babbush, Ryan and Benjamin, Simon C. and
             Endo, Suguru and Huggins, William J. and Li, Ying and
             McClean, Jarrod R. and O'Brien, Thomas E.},
  title   = {Quantum Error Mitigation},
  journal = {Reviews of Modern Physics},
  volume  = {95},
  number  = {4},
  pages   = {045005},
  year    = {2023}
}

@article{ng1999policy,
  author    = {Ng, Andrew Y. and Harada, Daishi and Russell, Stuart},
  title     = {Policy Invariance Under Reward Transformations:
               Theory and Application to Reward Shaping},
  booktitle = {Proc. 16th International Conference on Machine Learning (ICML)},
  pages     = {278--287},
  year      = {1999}
}

@inproceedings{wang2022quantumnat,
  author    = {Wang, Hanrui and Gu, Jiaqi and Ding, Yongshan and Li, Zirui and
               Chong, Frederic T. and Pan, David Z. and Han, Song},
  title     = {{QuantumNAT}: Quantum Noise-Aware Training with Noise Injection,
               Quantization and Normalization},
  booktitle = {Proc. 59th ACM/IEEE Design Automation Conference (DAC)},
  pages     = {1--6},
  year      = {2022}
}

@article{skolik2023robustness,
  author  = {Skolik, Andrea and Mangini, Stefano and B\"{a}ck, Thomas and
             Macchiavello, Chiara and Dunjko, Vedran},
  title   = {Robustness of Quantum Reinforcement Learning
             Under Hardware Errors},
  journal = {EPJ Quantum Technology},
  volume  = {10},
  number  = {1},
  pages   = {8},
  year    = {2023}
}

@article{ababei2011reconfiguration,
  author  = {Ababei, Cristinel and Kavasseri, Rajesh},
  title   = {Efficient Network Reconfiguration Using Minimum Cost Maximum
             Flow-Based Branch Exchanges and Random Walks-Based
             Loss Estimations},
  journal = {IEEE Transactions on Power Systems},
  volume  = {26},
  number  = {1},
  pages   = {30--37},
  year    = {2011}
}

@misc{ababei2009reds,
  author       = {Ababei, Cristinel},
  title        = {{REDS}: Repository of Distribution Systems},
  howpublished = {Online benchmark set},
  year         = {2009},
  note         = {Radial distribution test feeders for network
                  reconfiguration studies}
}

@ARTICLE{2020SciPy-NMeth,
  author  = {Virtanen, Pauli and Gommers, Ralf and Oliphant, Travis E. and
            Haberland, Matt and Reddy, Tyler and Cournapeau, David and
            Burovski, Evgeni and Peterson, Pearu and Weckesser, Warren and
            Bright, Jonathan and {van der Walt}, St{\'e}fan J. and
            Brett, Matthew and Wilson, Joshua and Millman, K. Jarrod and
            Mayorov, Nikolay and Nelson, Andrew R. J. and Jones, Eric and
            Kern, Robert and Larson, Eric and Carey, C J and
            Polat, {\.I}lhan and Feng, Yu and Moore, Eric W. and
            {VanderPlas}, Jake and Laxalde, Denis and Perktold, Josef and
            Cimrman, Robert and Henriksen, Ian and Quintero, E. A. and
            Harris, Charles R. and Archibald, Anne M. and
            Ribeiro, Ant{\^o}nio H. and Pedregosa, Fabian and
            {van Mulbregt}, Paul and {SciPy 1.0 Contributors}},
  title   = {{{SciPy} 1.0: Fundamental Algorithms for Scientific
            Computing in Python}},
  journal = {Nature Methods},
  year    = {2020},
  volume  = {17},
  pages   = {261--272},
  adsurl  = {https://rdcu.be/b08Wh},
  doi     = {10.1038/s41592-019-0686-2},
}
\clearpage

\section*{Appendix}
\label{sec:appendix}
\begin{table}[htbp]
    \centering
    \caption{IEEE 33 Bus Hyperparameters}
    \label{tab:mlp_33_params}
    \begin{tabular}{llc}
        \hline
        \textbf{Category} & \textbf{Hyperparameter} & \textbf{Value} \\
        \hline
        
        \multirow{5}{*}{Memory}
        & Priority utilization, $\alpha$ & 0.8 \\
        & Importance-sampling correction, $\beta$ & 0.35 \\
        & Minimum transition priority & $10^{-5}$ \\
        & Replay buffer capacity & 2500 \\
        & Bootstrap steps & 2500 \\
        
        \hline
        
        \multirow{13}{*}{Training}
        & Learning rate, $\eta$ & $4\times10^{-4}$ \\
        & Number of episodes & 2000 \\
        & Maximum episode length & 32 \\
        & Batch size & 128 \\
        & Learning rate decay & 0.01 \\
        & Discount factor, $\gamma$ & 0.90 \\
        & Initial exploration rate, $\epsilon_{\mathrm{init}}$ & 0.99 \\
        & Minimum exploration rate, $\epsilon_{\mathrm{min}}$ & 0.05 \\
        & Exploration decay & $0.4 \times$ episodes \\
        & Target-network update rate, $\tau$ & 0.006 \\
        & Target update interval & 100 \\
        & Evaluation interval & 10 \\
        & Q-value reduction & Mean \\
        
        \hline
        
        \multirow{5}{*}{Reward Parameters}
        & Voltage reward scale, $p_1$ & 6 \\
        & Voltage-deviation sensitivity, $p_2$ & 20 \\
        & Power-loss sensitivity, $p_3$ & 6 \\
        & Power-loss reward scale, $p_4$ & 1 \\
        & Topological reward scale, $p_5$ & 0.2 \\
        
        \hline
    \end{tabular}
\end{table}

\begin{table}[htbp]
    \centering
    \caption{IEEE 69 Bus Hyperparameters}
    \label{tab:mlp_69_params}
    \begin{tabular}{llc}
        \hline
        \textbf{Category} & \textbf{Hyperparameter} & \textbf{Value} \\
        \hline
        
        \multirow{5}{*}{Memory}
        & Priority utilization, $\alpha$ & 0.8 \\
        & Importance-sampling correction, $\beta$ & 0.35 \\
        & Minimum transition priority & $10^{-5}$ \\
        & Replay buffer capacity & 2500 \\
        & Bootstrap steps & 2500 \\
        
        \hline
        
        \multirow{13}{*}{Training}
        & Learning rate, $\eta$ & $4\times10^{-4}$ \\
        & Number of episodes & 2500 \\
        & Maximum episode length & 32 \\
        & Batch size & 256 \\
        & Learning rate decay & 0.01 \\
        & Discount factor, $\gamma$ & 0.92 \\
        & Initial exploration rate, $\epsilon_{\mathrm{init}}$ & 0.99 \\
        & Minimum exploration rate, $\epsilon_{\mathrm{min}}$ & 0.05 \\
        & Exploration decay & $0.4 \times$ episodes \\
        & Target-network update rate, $\tau$ & 0.006 \\
        & Target update interval & 100 \\
        & Evaluation interval & 10 \\
        & Q-value reduction & Mean \\
        
        \hline
        
        \multirow{5}{*}{Reward Parameters}
        & Voltage reward scale, $p_1$ & 6 \\
        & Voltage-deviation sensitivity, $p_2$ & 20 \\
        & Power-loss sensitivity, $p_3$ & 6 \\
        & Power-loss reward scale, $p_4$ & 1 \\
        & Topological reward scale, $p_5$ & 0.2 \\
        
        \hline
    \end{tabular}
\end{table}

\end{document}